\documentclass{aastex}

\usepackage{spr-astr-addons}
\usepackage{url}

\begin{document}

\title{New local interstellar spectra for protons, Helium and Carbon derived from PAMELA and Voyager 1 observations}

\shorttitle{New LIS derived from V1 and PAMELA}
\shortauthors{Bisschoff & Potgieter}

\author{D. Bisschoff}
\and
\author{M.S. Potgieter}
\affil{Centre for Space Research, North-West University, 2520 Potchefstroom, South Africa; tel: +27182994016; driaan.b@gmail.com}

\begin{abstract}

With the cosmic ray observations made by the Voyager 1 spacecraft outside the dominant modulating influence of the heliosphere, the comparison of computed galactic spectra with experimental data at lower energies is finally possible. Spectra for specifically protons, Helium and Carbon nuclei, computed by galactic propagation models, can now be compared with observations at low energies from Voyager 1 and at high energies from the PAMELA space detector at Earth. We set out to reproduce the Voyager 1 observations in the energy range of 6 MeV/nuc to 60 MeV/nuc, and the PAMELA spectrum above 50 GeV/nuc, using the GALPROP code, similarly to our previous study for Voyager 1 electrons. By varying the galactic diffusion parameters in the GALPROP plain diffusion model, specifically the rigidity dependence of spatial diffusion, and then including reacceleration, we compute spectra simultaneously for galactic protons, Helium and Carbon.We present new local interstellar spectra, with expressions for the energy range of 3 MeV/nuc to 100 GeV/nuc, which should be of value for solar modulation modeling.
\end{abstract}


\section{Introduction}

In August 2012, Voyager 1 (V1) crossed the heliopause (HP) at a distance of 121.7\,AU and then began to measure cosmic rays (CRs) outside the dominant influence of the heliosphere for the first time \citep{Stone2013}. These observations for specifically galactic protons, Helium and Carbon, allow the comparison of computed galactic spectra with experimental data down to a few MeV/nucleon (MeV/nuc). With the addition of high energy observations made in low earth orbit by PAMELA \citep{Pamela2011protonhelium,Pamela2014boroncarbon,Menn2013,Boezio2014}, estimations of the local interstellar spectra (LIS's) over a very wide range of energies can be made more reliably than done previously when making use of a comprehensive galactic propagation model, such as the GALPROP code \citep{StrongReview2007,Webrun,Moskalenko}. 

We aim to reproduce the V1 observations for CR protons, Helium ($^3$He$_2$ + $^4$He$_2$) and Carbon ($^{12}$C$_6$ + $^{13}$C$_6$), while also matching the PAMELA observations for all three CR species. This study endeavors to present a set of LIS's, with expressions, for these three CR species that can be used further in other CR study fields, especially for heliospheric modulation studies \citep{PotgieterReview2013} where these LIS's are used as input spectra, serving as initial conditions. We attempt to achieve this by using the GALPROP propagation code in its simplest form over the energy range 3 MeV/nuc to 100 GeV/nuc. These LIS's are calculated by varying the galactic diffusion parameters in the model and including additional features as needed, such as adding reacceleration and adjusting the source input. By eliminating LIS's that do not agree with the required observational restrictions, we attempt to find a single set of parameters to reproduce the mentioned observed spectra simultaneously. This is done similarly to the modeling we used to achieve this for galactic electrons \citep{Bisschoff2014}. Empirically derived LIS's based on V1 and PAMELA observations above energies where solar modulation becomes negligible \citep[e.g.][]{StraussPotgieter2014} were reported before by \citet{Potgieter2014} and \citet{Potgieter2014Conf}. For a complete description of solar modulation effects for protons related to PAMELA observations in 2009, see \cite{Vos2013} and \cite{Potgieter2014Solar}.

\section{The numerical model and assumptions}

The cosmic ray equation for galactic propagation generally has the form:
\begin{equation}
\label{eqn:1}
\begin{aligned}
\frac{\partial \psi}{\partial t} & = S(\mathbf{r},p)+\nabla\cdot(K\nabla\psi-\mathbf{V}\psi) \\
& + \frac{\partial}{\partial p}\left[p^2 K_{p}\frac{\partial}{\partial p}\frac{1}{p^2}\psi + \frac{p}{3}(\nabla\cdot\mathbf{V})\psi - \dot{p}\psi\right] \\
& - \frac{1}{\tau_{f}}\psi-\frac{1}{\tau_{r}}\psi,
\end{aligned}
\end{equation}
where $\psi = \psi(\mathbf{r},p,t)$ is the density per unit of total particle momentum, $S(\mathbf{r},p)$ is the source term, $K$ is the spatial diffusion coefficient, $\mathbf{V}$ is the convection velocity, reacceleration is described as diffusion in momentum space and determined by the coefficient $K_{p}$, $\dot{p}$ is the momentum loss rate, $\tau_f$ is the timescale for fragmentation and depends on the total spallation cross-section and $\tau_r$ the timescale for radioactive decay. For details on the basic theory and concepts of cosmic ray propagation in the Galaxy, see the review by \cite{StrongReview2007}.

For cosmic ray propagation studies, the Galaxy is usually described as a cylindrical disk with a radius of $\sim 20$ kpc and a height of up to $\sim 4$ kpc, including the galactic halo, in which cosmic rays have a finite chance to return to the galactic disk. Assuming symmetry in azimuth leads to two spatial dimensional (2D) models that depend simply on galactocentric radius and height, whereas neglecting time dependence leads to steady-state models. As mentioned, we decided on using a plain diffusion approach initially, to keep the modeling as simple as possible. When implemented in the GALPROP code, it gives a 2D model with radius $r$, the halo height $z$ above the galactic plane and symmetry in the angular dimension in galactocentric-cylindrical coordinates. The halo height was fixed to $z = 4$ kpc and was kept constant because varying its size can simply be counteracted by directly varying the diffusion coefficient. In this plain diffusion model, the velocity and gradient in the galactic wind is set to zero. When considering reacceleration the momentum-space diffusion coefficient $K_{p}$ is estimated as related to $K$ so that $K_{p} K\propto p^2 v_A ^2$, with $v_A$ the Alfve\'n wave speed set to 36 km\,s$^{-1}$. Reacceleration is not considered for a plain diffusion model and for this case $v_A = 0$. Other parameters in the model (such as source abundance values, interstellar properties, cross sections and gas densities) were also investigated, but for this study they were adapted straightforwardly from \citet{Ptuskin2006} and are not repeated here.

In this simplified approach, the spatial diffusion coefficient is assumed to be independent of $r$ and $z$. It is taken as being proportional to a power-law in rigidity $P$ so that:
\begin{equation}
\label{eqn:2}
K=\beta K_{0}(P /P_{0})^{\delta},
\end{equation}
where $\delta=\delta_{1}$ for rigidity $P < P_{0}$ (the reference rigidity), $\delta=\delta_{2}$ for $P > P_{0}$ and with $\beta = v/c$ the dimensionless particle velocity given by the speed of particles $v$, at a given rigidity relative to the speed of light, $c$. Here, $K_0$ is the scaling factor for diffusion in units of 10$^{28}$\,cm$^{2}$\,s$^{-1}$.

The injection spectrum for nuclei, as input to the source term, is assumed to be a power-law in rigidity so that: 
\begin{equation}
\label{eqn:3}
S(P) \propto (P /P_{\alpha 0})^{\alpha}, 
\end{equation}
for the injected particle density and usually contains a break in the power-law with indices $\alpha _1$ and $\alpha _2$ below and above the source reference rigidity $P_{\alpha 0}$, respectively. Values for $\alpha _1$ and $\alpha _2$ are positive and non-zero, thus giving a rigidity dependent injection spectrum. Only primary nuclei are given an input spectrum, isotopes considered wholly secondary are set to 0 at the sources. See also \citet{StrongMoskalenko1998}. Source abundance values were generally kept unchanged from \citet{Ptuskin2006}, except for $^4$He$_2$ which we assigned a relative increase, as shown and discussed in Section \ref{S:3}.

In what follows, we show several proton, Helium and Carbon LIS's computed with the GALPROP code which solves the given transport equation using a Crank-Nicholson implicit second-order scheme. These computational runs were done via the GALPROP WebRun service (\url{http://galprop.stanford.edu/webrun/}) \citep{Webrun}. A description of the GALPROP model, and the theory it is based on, can be found in the overviews by \cite{StrongReview2007} and \cite{Moskalenko} and references therein; see also the GALPROP Explanatory Supplement available from the GALPROP website.

\section{Results: Reproducing the Voyager 1 proton observations beyond the HP with a plain diffusion model}
\label{S:3}
\begin{figure}[tb]
  \centering
  \includegraphics[width=\columnwidth]{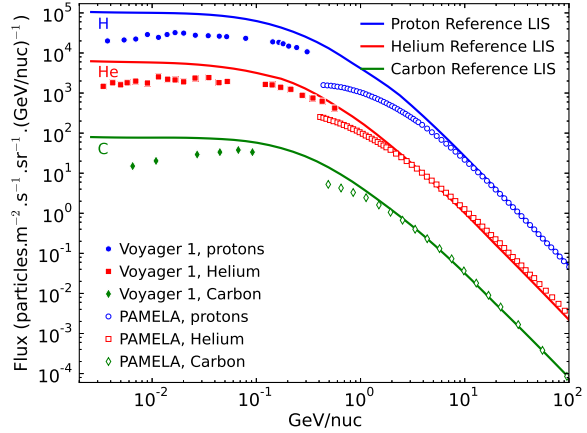}
  \caption{Reference model LIS's for protons (blue curve), Helium (red curve) and Carbon (green curve) shown alongside the observational data from the V1 spacecraft at 121.7\,AU \citep{Stone2013} (closed symbols) and modulated spectra from PAMELA at Earth, averaged for 2006-2008  \citep{Pamela2011protonhelium,Pamela2014boroncarbon,Menn2013} (open symbols). Note that solar modulation gets increasingly larger below about 20 GeV}
  \label{fig:1}
\end{figure}

When comparing the V1 observations for protons, Helium and Carbon \citep{Stone2013} to the LIS's computed by previous propagation models, it becomes clear that the models mostly overestimated the intensity of the LIS's below about 1 GeV/nuc.\cite[See e.g.][for an evaluation on LIS's before V1 crossed the HP]{Herbst2012}. This is illustrated first in Figure \ref{fig:1}, where the V1 observations are compared to such LIS's produced with the GALPROP code, which we consider to be our reference model LIS's for protons (blue), Helium (red) and Carbon (green). These observations clearly show that any further estimations of the LIS's produced by numerical models would have to reduce the spectral intensity for the observed V1 energy range.

To achieve this we continue with the GALPROP model in its simplest form, the 2D plain diffusion model. The parameters $K_0$, $P_0$ and $\delta _1$ from Eq. \ref{eqn:2} determine the rigidity dependence of the diffusion coefficient and are basically considered as free parameters for this study. The parameter $\delta _2$ was initially also taken as such, but investigative tests showed that the reference value of $\delta _2 = 0.6$ is needed in order for the LIS's to reproduce the observations. The value of $\delta _2$ was thus kept unchanged for all the plain diffusion model runs. For the source function the values required in Eq. \ref{eqn:3} were kept fixed at $P_{\alpha 0} = 40$ GV, $\alpha _1 = 2.30$ and $\alpha _2 = 2.15$.

\begin{figure}[tb]
  \centering
  \includegraphics[width=\columnwidth, trim=10 0 46 33,clip=true,draft=false]{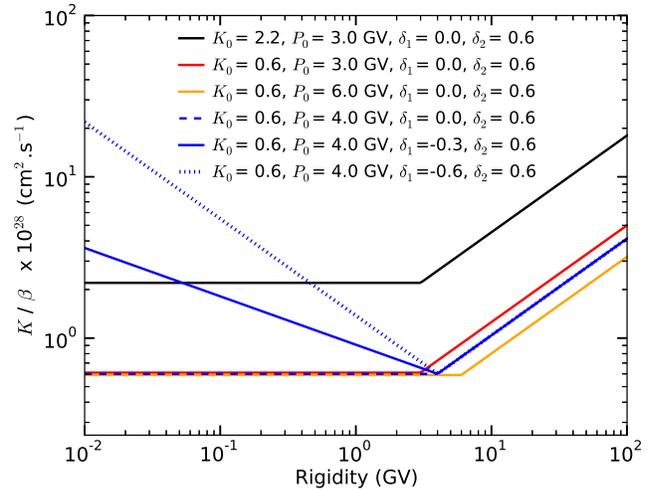}
  \caption{The assumed diffusion coefficient, as function of rigidity, as implemented for the reference model of Fig. \ref{fig:1} is shown in black. Adjusted model parameters are chosen as to potentially reproduce the V1 proton observations. This gives the listed values for $K_0$, $P_0$ and $\delta _1$, and gives the diffusion coefficients shown with the red, yellow and three blue lines}
  \label{fig:2}
\end{figure}

In order to simply reproduce the observations, $K_0$ and $P_0$ were adjusted together, with separate model runs adjusting $\delta _1$. This manner of choosing parameter values showed that only adjusting $\delta _1$ could not achieve the reproduction of the V1 observations, although $\delta _1$ does affect the shape of the computed LIS's significantly in the required energy range. After finding the sets of values for $K_0$ and $P_0$ that most closely give computed LIS's that reproduce the V1 data, $\delta _1$ was then adjusted to finely control the computed LIS shape. The resulting sets of assumed diffusion coefficients, called models, are shown in Figure \ref{fig:2}, together with the parameters used for the reference spectra shown in Figure \ref{fig:1}.

\begin{figure}[tb]
  \centering
  \includegraphics[width=\columnwidth, trim=10 0 46 33,clip=true,draft=false]{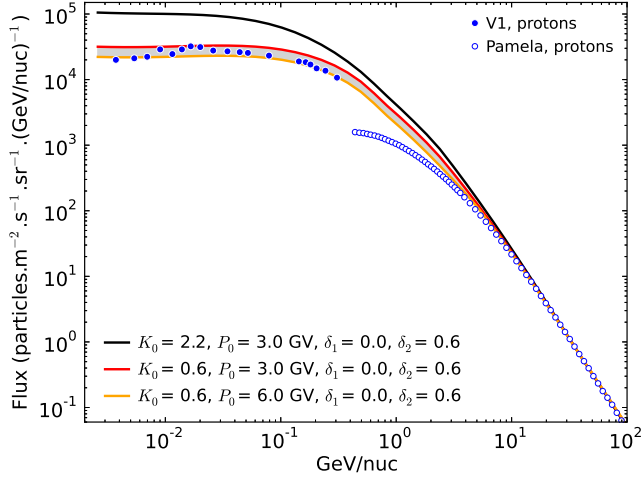}
  \caption{The computed proton LIS band bounded by the parameters listed in Fig. \ref{fig:2}, upper value given by the red curve and lower value by the yellow curve. These two LIS's, together with the reference LIS (black curve), are compared to the V1 and PAMELA proton data. A computed LIS reproducing the V1 data should ideally lie inside this band, given the spread in the observations. The PAMELA observations above 20 GeV are clearly reproduced well where solar modulation becomes negligible}
  \label{fig:3}
\end{figure}

Adjusting only $K_0$ and $P_0$ gives the models with $K_0 = 6.0 \times 10^{27}$ cm$^2$\,s$^{-1}$ , $P_0 = 3.0$ GV (red line) and $K_0 = 6.0 \times 10^{27}$ cm$^2$\,s$^{-1}$ , $P_0 = 6.0$ GV (orange line). These models produce the LIS band in Figure \ref{fig:3}, in comparison with the computed LIS of the reference model (black line). These computed LIS's give a lower (red curve) and upper (orange curve) value of the proton LIS needed to match the V1 spectrum. A computed proton LIS reproducing the V1 proton data should ideally lie inside this band, given the spread in the observations. The higher value of $P_0$ (orange curve) results in a lower diffusion coefficient giving a lower intensity for the LIS at energies below 10 GeV/nuc.  This band represents the most reasonable computed LIS when considering diffusion coefficients where the indices are the same as those of the reference model.

\begin{figure}[tb]
  \centering
  \includegraphics[width=\columnwidth, trim=10 0 46 33,clip=true,draft=false]{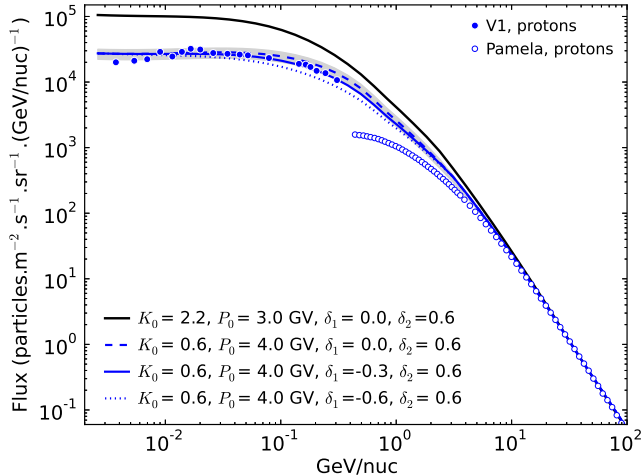}
  \caption{Computed proton LIS's for the remaining models in Fig. \ref{fig:2} (three blue curves) compared to the band (shaded grey) of Fig. \ref{fig:3} and the reference LIS (black line), showing the change in slope for the LIS with a decrease in $\delta _1$. For $\delta _1 = -0.3$ the LIS satisfactorily matches the V1 protons, while only overestimating the values below 7 MeV/nuc}
  \label{fig:4}
\end{figure}

Attempting to give a better representation of the observations than the LIS band of Figure \ref{fig:3}, parameters are chosen to give models as shown by the three blue lines in Figure \ref{fig:2}. The value of $K_0$ is kept the same, but $P_0$ is now set to 4.0 GV and for two of these models $\delta_1$ is adjusted to $-0.3$ (solid blue line) and $-0.6$ (dotted blue line) instead of 0. Figure \ref{fig:4} shows the three corresponding computed LIS's. The change in $\delta_1$ greatly increases the diffusion coefficient at the lowest rigidities and evidently lower the computed LIS's for energies below about 4 GeV/nuc. Effectively the LIS shape is changed, creating a sharper change in the LIS's at about 4 GeV/nuc and a flatter shape below this spectral change. The resulting proton LIS for $\delta _1 = 0$ (dashed blue curve) agrees well with the data only at lower V1 energies, while for $\delta_1 = -0.6$ the LIS only matches the highest energy observations of V1. The resulting proton LIS for $\delta_1 = -0.3$ reproduces the V1 data well above 7 MeV/nuc and is the best representation of the observations within the band. 

\begin{figure}[tb]
  \centering
  \includegraphics[width=\columnwidth, trim=10 0 46 33,clip=true,draft=false]{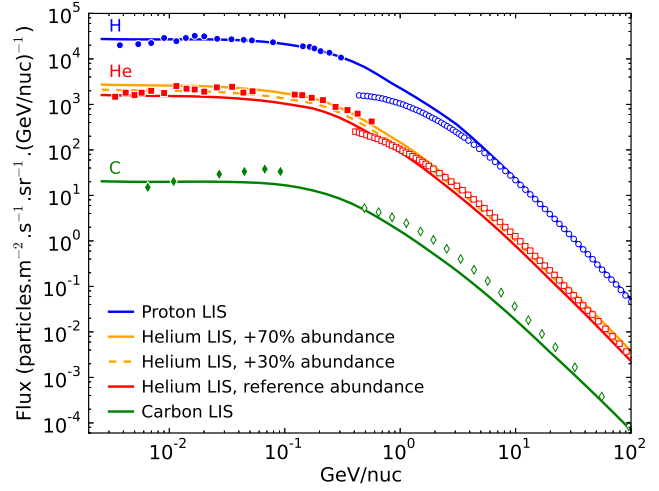}
  \caption{Matching computed proton LIS (blue curve) from Fig. \ref{fig:4}, the corresponding computed Helium LIS (red curve) and computed Carbon LIS (green curve) compared to the observational data as in Fig. \ref{fig:1}. The computed LIS for Helium with an increased $^4$He$_2$ source abundance is shown for an increase of 30\% (dashed yellow curve) and an increase of 70\% (solid yellow curve)}
  \label{fig:5}
\end{figure}

With the computed proton LIS reproducing the V1 proton observations, our attention turns to the Helium and Carbon LIS's. Figure \ref{fig:5} shows the LIS's computed with $K_0 = 6.0 \times 10^{27}$, $P_0 = 4.0$ GV and $\delta _1 = -0.3$, for protons (blue curve), Helium (red curve) and Carbon (green curve). Unfortunately, this computed Helium LIS is lower than the V1 values, as well as having a lower intensity than the PAMELA observations above 4 GeV/nuc, similarly to the reference Helium LIS in Figure \ref{fig:1}. To correct this discrepancy, while keeping the proton LIS unchanged, the relative abundance of $^4$He$_2$ (the primary Helium isotope at the sources) needs to be increased. To match the PAMELA values above 50 GeV/nuc, where heliospheric modulation can be safely ignored \citep{StraussPotgieter2014}, an increase of about 30\% in the $^4$He$_2$ abundance is sufficient, as is represented in Figure \ref{fig:5} with the dashed orange curve. As expected the proton and Carbon LIS's is unaffected, while the Helium LIS has an increased intensity over all energies and now matches the PAMELA values above 50 GeV/nuc. For the V1 data above 0.1 GeV/nuc this Helium LIS remains too low, while also undercutting the PAMELA data at energies below 50 GeV/nuc, which from a solar modulation point of view is not suitable. The effect of a 70\% increase to the $^4$He$_2$ abundance is also shown in Figure \ref{fig:5} with the solid orange curve. This computed Helium LIS achieves the required effect as the LIS matches the V1 data well (above 0.01 GeV/nuc) while keeping above the PAMELA data at all energies, which is a requirement for a solar modulation; an assumed or computed LIS cannot be lower than the observed spectrum at the Earth. The drawback is that the computed LIS being slightly higher than the PAMELA values at energies above 50 GeV/nuc where it is expected to match more closely.

The computed Carbon LIS shown in Figure \ref{fig:5} gives a relatively poor representation of the measured intensities. Increasing the Carbon abundances similarly to the Helium abundances could improve the LIS over most energies, but the slope of the LIS then resulted in the PAMELA observations above 10 GeV/nuc not being matched, while the reference LIS of Figure \ref{fig:1} matches the slope suggested by these observations. The corresponding computed B/C ratio, as shown in Figure \ref{fig:10}, also then greatly overestimates the values over all energies, while the reference model gives a good representation of the observed ratio, but evidently fails to reproduce the V1 spectra. This indicates that simply changing the parameters of the plain diffusion model is insufficient to compute acceptable LIS's for protons, Helium and Carbon simultaneously.

\section{Results: Including reacceleration in the propagation model}

Addressing the above mentioned problem, we turned to the reacceleration model as implemented in GALPROP. This model is less simplistic as it also includes diffusion in momentum space. The variables for the reacceleration propagation model are similar to that of the plain diffusion model in GALPROP, but now includes an Alfv\'en wave speed $v_A = 36$ km\,s$^{-1}$ and particularly a single index $\delta$ in Eq. \ref{eqn:2}, above and below the reference rigidity. Consequently , a new reference model is derived from the approach of \citet{Ptuskin2006} that included reacceleration. The source function parameters in Eq. \ref{eqn:3} were also changed correspondingly with $P_{\alpha 0} = 9.0$ GV, $\alpha _1 = 1.82$ and $\alpha _2 = 2.36$.

\begin{figure}[tb]
  \centering
  \includegraphics[width=\columnwidth, trim=10 0 46 33,clip=true,draft=false]{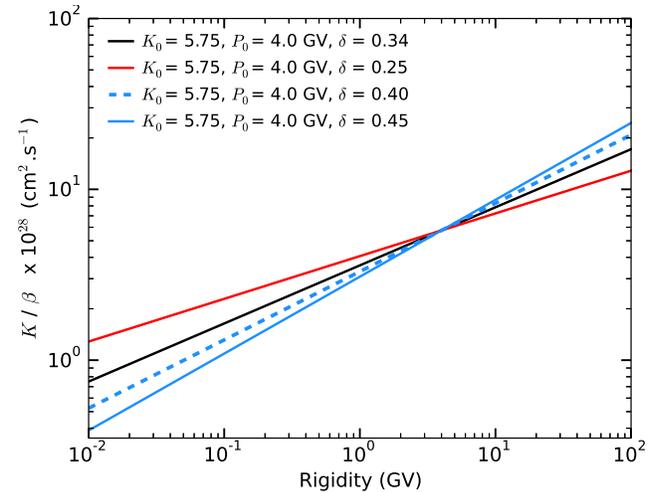}
  \caption{The chosen parameter values for the reacceleration model, with variation in $\delta$: 0.25 (solid blue line), 0.30 (dashed blue line), 0.45 (solid red line), and the reacceleration reference model with index of 0.34 (black line). $K_0$ is kept at $5.75\times 10^{28}$ cm$^2$\,s$^{-1}$ and $P_0$ at 4.0 GV}
  \label{fig:6}
\end{figure}

\begin{figure}[tb]
  \centering
  \includegraphics[width=\columnwidth, trim=10 0 46 33,clip=true,draft=false]{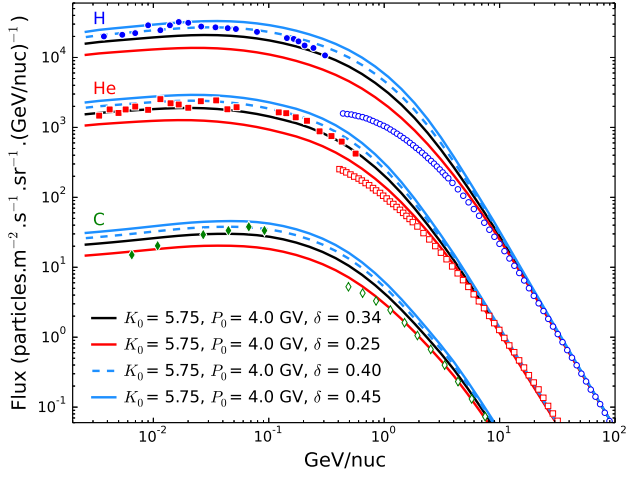}
  \caption{Computed LIS's for the parameters given in Figure \ref{fig:6} (with the same curve colour coding) with protons at the top, Helium in the middle and Carbon at the bottom. These LIS's are compared to the observational data as in Fig. \ref{fig:1}. Models with a larger $\delta$ evidently give LIS's with increased intensity to better match the V1 data than the reference LIS}
  \label{fig:7}
\end{figure}

To reproduce the measured CRs intensities, the diffusion coefficient parameters of the new reference model are adjusted similarly to what was done for the plain diffusion model. The index $\delta$ is initially adjusted while keeping the values of $K_0$ and $P_0$ the same. The parameters are shown in Figure \ref{fig:6} where $\delta$ is adjusted to 0.25 (red line), 0.40 (dashed blue line) and 0.45 (solid blue line) from the reacceleration reference index of 0.34 (black line), with $K_0 = 5.75\times 10^{28}$ cm$^2$\,s$^{-1}$ and $P_0 = 4.0$ GV. The corresponding computed LIS's for these parameter values are shown in Figure \ref{fig:7}.  Lowering the value of $\delta$ decreases the computed LIS intensity, as shown for $\delta$ = 0.25, a steeper index than the reference value is clearly needed. For $\delta = 0.45$ the computed LIS's (solid blue curves) for protons, Helium and Carbon give upper values to the CR spectra in the V1 energy range, while $\delta = 0.25$ LIS (red curve) gives the lower values. The computed LIS for $\delta$ = 0.40 match both the V1 and PAMELA observations well, but none of the computed Carbon LIS's can reproduce the decreasing trend suggested by the lowest energy of V1 observations. These computed LIS's show higher intensities in the energy range 0.1 GeV/nuc to 10 GeV/nuc for protons when compared to the plain diffusion LIS of Figure \ref{fig:5}. In this energy range solar modulation should be considered when comparing the LIS's to the PAMELA observations, the plain diffusion LIS lies too close to the observations, seemingly underestimating the amount of expected modulation when compared to the estimations of \citet{StraussPotgieter2014}. Including reacceleration in the model gives a LIS that more closely reproduces the expected modulation.

\begin{figure}[tb]
  \centering
  \includegraphics[width=\columnwidth, trim=10 0 46 33,clip=true,draft=false]{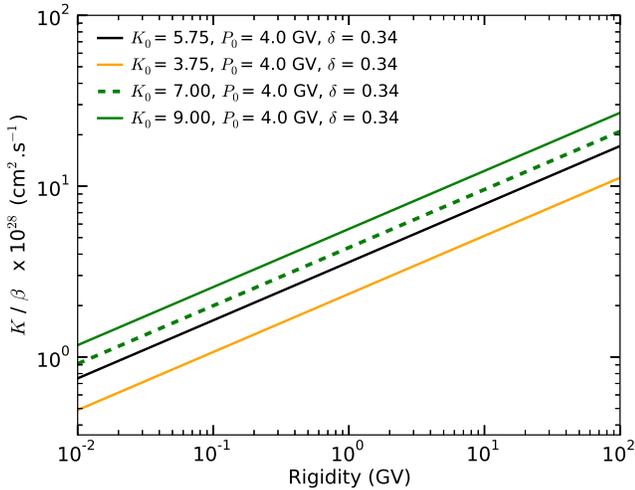}
  \caption{The newly chosen parameter values for reacceleration models, with variation in $K_0$: 3.75$\times 10^{28}$ cm$^2$\,s$^{-1}$ (solid yellow line), 7.00$\times 10^{28}$ cm$^2$\,s$^{-1}$ (dashed green line), 9.00$\times 10^{28}$ cm$^2$\,s$^{-1}$(solid green line), compared with the reference model with a value of 5.75$\times 10^{28}$ cm$^2$\,s$^{-1}$ (dashed black line). $P_0$ is kept at 4.0 GV and $\delta$ at 0.34}
  \label{fig:8}
\end{figure}

\begin{figure}[tb]
  \centering
  \includegraphics[width=\columnwidth, trim=10 0 46 33,clip=true,draft=false]{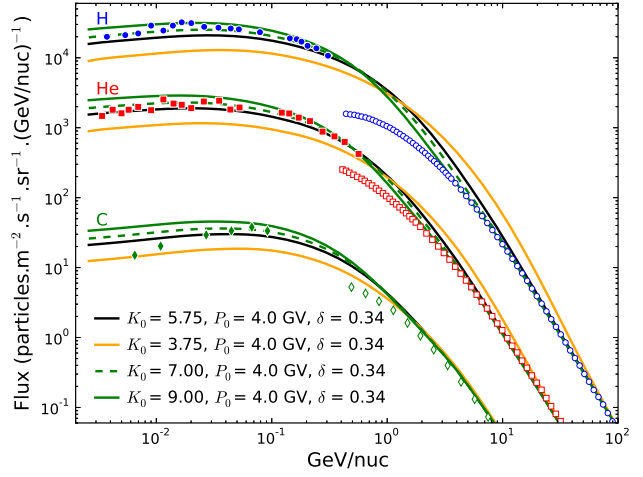}
  \caption{Computed LIS's for the parameters given in Fig. \ref{fig:8} with protons at the top, Helium in the middle and Carbon at the bottom. The models with a lower $K_0$ give lower intensities below 1 GeV/nuc, but give higher intensities above this energy. Higher values of $K_0$ work in the opposite direction.}
  \label{fig:9}
\end{figure}

\begin{figure}[tb]
  \centering
  \includegraphics[width=\columnwidth, trim=30 0 46 33,clip=true,draft=false]{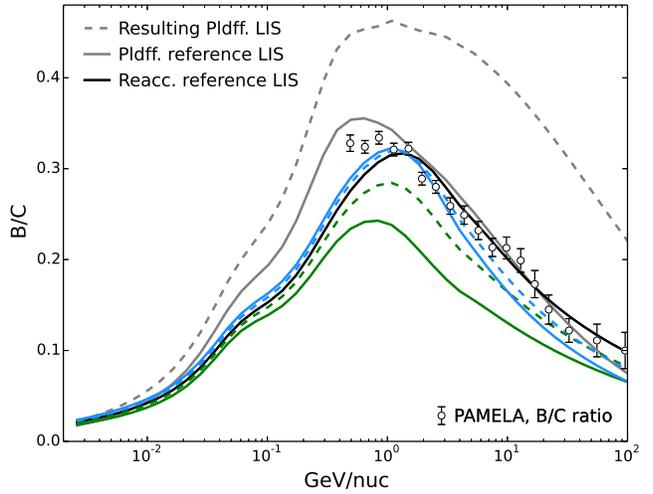}
  \caption{Computed B/C ratios for the most suitable models compared to the PAMELA B/C observations (open circles). The plain diffusion model (Pldff) of Fig. \ref{fig:5} (dashed grey curve) overestimates the observed values over all energies, while the reacceleration models reproduce the ratio more closely. The reference models match the B/C ratio well above 1 GeV/nuc. From our reacceleration test runs the model with $\delta$ = 0.40 (dashed blue curve) shows the better match, with the other promising models from Fig. \ref{fig:7} (solid blue curve) and Fig. \ref{fig:9} (solid green curve and dashed green curve) underestimated the ratio}
  \label{fig:10}
\end{figure}

The next set of reacceleration models had only the $K_0$ values adjusted, because a break in the index $\delta$ is not included, the value of $P_0$ could also be adjusted with the same effect, but is kept the same for these runs. The models are shown in Figure \ref{fig:8} where $K_0$ is adjusted to $3.75\times 10^{28}$ cm$^2$\,s$^{-1}$ (solid yellow line), $7.00\times 10^{28}$ cm$^2$\,s$^{-1}$ (dashed green line), and $9.00\times 10^{28}$ cm$^2$\,s$^{-1}$ (solid green line) from the reference value of $5.75 \times 10^{28}$ cm$^2$\,s$^{-1}$ (black line), while $P_0$ is kept as 4.0 GV and $\delta = 0.34$. Figure \ref{fig:9} shows the corresponding computed LIS's resulting from these adjustments. For a lower $K_0$ the LIS intensity is decreased at lower energies, while increasing intensity for energies above about 1 GeV/nuc as follows for $K_0 = 3.75\times 10^{28}$ cm$^2$\,s$^{-1}$. The inverse is true for a higher $K_0$, giving increased intensity at lower energies and decreased intensity for $E >$ 1 GeV/nuc. This decrease is too large for $K_0 = 9.00\times 10^{28}$ cm$^2$\,s$^{-1}$ as it matches the the PAMELA values closely above 3 GeV/nuc, thus ignoring requirements of heliospheric modulation completely. For $K_0 = 7.00\times 10^{28}$ cm$^2$\,s$^{-1}$ the decrease is less, while giving a good match to the V1 observations over all energies for protons and Helium. However, once again the steeper trend of the V1 Carbon below 60 MeV/nuc can't be reproduced.

The most suitable models, from our point of view, are selected from all the above tests and the computed  B/C ratios for these models are shown in Figure \ref{fig:10} in comparison with the PAMELA B/C observations from \citet{Pamela2014boroncarbon}. This includes the reacceleration models with $\delta$ = 0.40 (dashed blue curve) and 0.45 (solid blue curve) of Figure \ref{fig:6} and the models with $K_0 = 7.00 \times 10^{28}$ cm$^2$\,s$^{-1}$ (dashed green line) and 9.00$\times 10^{28}$ cm$^2$\,s$^{-1}$ (solid green line) of Figure \ref{fig:8}. The plain diffusion model from Figure \ref{fig:2} with $\delta_1 = -0.3$, $K_0 = 6.0 \times 10^{27}$ cm$^2$\,s$^{-1}$ and $P_0 = 4.0$ GV is also shown (dashed grey curve). Both the reference models, for plain diffusion (grey curve) and reacceleration (black curve), match the PAMELA B/C values quite well. 

As stated in the previous section, our best plain diffusion model does not reproduce the observed ratio and lies well outside the uncertainty range. In contrast the  reacceleration models produce ratios that better match the PAMELA values. The two $K_0$ adjusted models underestimate the ratio, with $K_0 = 7.00\times 10^{28}$ cm$^2$\,s$^{-1}$ (dashed green line) giving a reasonable match at high energies. The two $\delta$ adjusted models matching the ratio well for energies above 1 GeV/nuc, but $\delta = 0.45$ (solid blue curve) underestimating slightly at higher energies. Of our four reacceleration models $\delta = 0.40$ (dashed blue curve) gives the most statisfactorily reproduction of the B/C ratio, with only the lowest energy PAMELA observations not reproduced.

Using these B/C ratios the other models can be eliminated to arrive at our prefered model parameter set: $K_0 = 5.75 \times 10^{28}$ cm$^2$\,s$^{-1}$, $P_0 = 4.0$ GV and $\delta =$ 0.40, with no breaks in the rigidity dependence of the diffusion coefficient in these GALPROP models.

\section{Discussion and Conclusions}

\begin{figure}[bt]
  \centering
  \includegraphics[width=\columnwidth, trim=10 0 46 33,clip=true,draft=false]{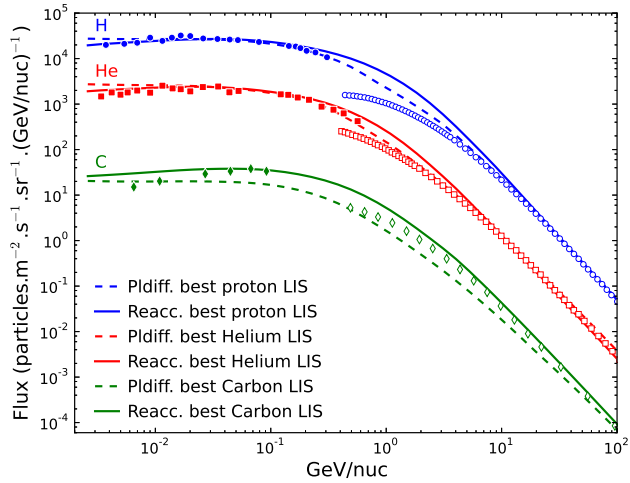}
  \caption{The self-consistent LIS's computed in this study to reproduce the measured CR spectra for V1 and PAMELA. Protons are shown in blue, Helium in red and Carbon in green. The dashed curves were produced with the plain diffusion model (indicated as Pldiff) and match the proton and Helium data well, but not the Carbon. The solid curves are from the reacceleration model (indicated as Reacc.) and evidently reproduce the data sets, especially the Carbon data, satisfactorily}
  \label{fig:11}
\end{figure}

By adjusting the diffusion coefficient in the GALPROP code over a large enough parameter space, the effect of the diffusion parameters $K_0$, $P_0$ and $\delta _1$ on the computed proton, Helium and Carbon LIS's, could be investigated. With this knowledge the intensity of the computed LIS's was decreased from that of the plain diffusion reference model to those matching the V1 proton and Helium observations made outside the heliosphere. The parameters $K_0$, $P_0$ and the index $\delta_1$ required adjusting (and the source abundance in the case of $^4$He$_2$) in the plain diffusion model. To improve on the relatively poor reproducing of the Carbon LIS and B/C ratio with the plain diffusion model, reacceleration was included. When considering this reacceleration approach, only adjusting the index $\delta$ is required to match the observational data. This resulted in the computed LIS's for protons, Helium and Carbon shown by the solid lines in Fig. \ref{fig:11} in comparison with the solutions of the plain 
diffusion model (dashed lines).

The computed LIS's produced by the plain diffusion model reproduce the V1 data over all energies for both protons and Helium, except for the values below 0.01 GeV/nuc. This decreasing intensity trend for CR intensities at lower energies is present in both the proton and Helium observations and not seen in any LIS's produced with the plain diffusion model, but can be seen in the reacceleration model LIS's. This suggests that reacceleration is required to reproduce this feature in the observed spectra. The computed LIS's for the reacceleration models reproduce the V1 data quite reasonably well over all energies, as well as reproducing the Carbon observations. Only the values for protons and Helium above 0.2 GeV/nuc are better represented via the plain diffusion model. None of the considered models could reproduce the steeper decreasing trend seen in the V1 Carbon observations below 60 MeV/nuc.

For the PAMELA data, the plain diffusion proton LIS matches the observations closely above 5 GeV/nuc, but the reacceleration model suggests an amount of modulation in this energy range that more closely follows the estimations of \citet{StraussPotgieter2014}. Our plain diffusion Helium LIS, with an increased source abundance, is higher than the PAMELA values at energies above 50 GeV/nuc. While the LIS is expected to match the observed values more closely, as the heliospheric modulation is negligible in this region, it is an improvement on the plain diffusion reference model and this computed LIS still meets the other requirements. The reacceleration Helium LIS also shows an acceptable amount of modulation below 10 GeV/nuc, but above this energy slightly underestimates the PAMELA values, this is not easily rectified without influencing the lower energy data reproduction. To determine if any of these LIS's really reflect the amount of modulation in the heliosphere, the LIS's would have to be examined closer using an advanced heliospheric modulation code \citep[e.g.][]{Potgieter2014Solar}, not just a simple force-field modulation approach. The reacceleration model greatly improves on estimating the Carbon observations, specifically the high energy spectral slope and overall intensity, neither of which the plain diffusion model could reproduce satisfactorily. 

As with our previous study where we presented an electron LIS to match the V1 observations \citep{Bisschoff2014}, we have shown here LIS's for both protons and Helium that reproduce the V1 observations outside the heliosphere, with both a plain diffusion and a reacceleration model, while still adhering to the limits set forth by the PAMELA observations at Earth. The plain diffusion model overestimates the B/C ratios and does not reproduce the Carbon observations when the diffusion parameters are adjusted to decrease the proton intensity. We therefore prefer the GALPROP based reacceleration model to reproduce the V1 observations and find the model with $P_{\alpha 0}$ = 9.0 GV, $\alpha _1 = 1.82$, $\alpha _2 = 2.36$, $v_A = 36$ km\,s$^{-1}$, $K_0 = 5.75 \times 10^{28}$ cm$^2$\,s$^{-1}$, $P_0 = 4.0$ GV, $\delta = 0.40$ to be the best representation.

\begin{figure}[tb]
  \centering
  \includegraphics[width=\columnwidth, trim=10 0 46 33,clip=true,draft=false]{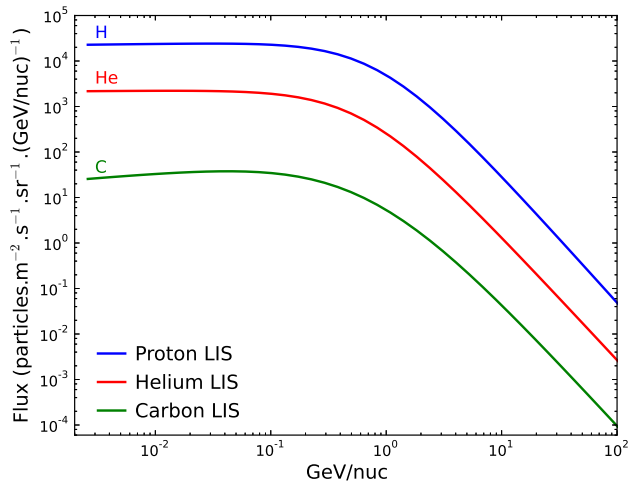}
  \caption{The mathematically approximated LIS's based on the computed LIS's as shown in Fig. \ref{fig:11}: For protons by Eq. \ref{eqn:4} (blue curve), for Helium by Eq. \ref{eqn:5} (red curve) and for Carbon by Eq. \ref{eqn:6} (green curve)}
  \label{fig:12}
\end{figure}

The computed LIS's resulting from these parameters can be approximated (within 12\%) over the energy range 3 MeV/nuc to 100 GeV/nuc by the following expressions. The approximate proton LIS is given by:
\begin{equation}
\label{eqn:4}
 J_{\rm p}(E) = 3719.0 \, \frac{1}{\beta ^2} \, E ^{1.03} \left(\frac{E^{1.21} + 0.77^{1.21}}{1 + 0.77^{1.21}} \right) ^{-3.18},
\end{equation}
the approximate Helium LIS is given by:
\begin{equation}
\label{eqn:5}
 J_{\rm He}(E) = 195.4 \, \frac{1}{\beta ^2} \, E ^{1.02} \left(\frac{E^{1.19} + 0.60^{1.19}}{1 + 0.60^{1.19}} \right) ^{-3.15},
\end{equation}
and the approximate Carbon LIS is given by:
\begin{equation}
\label{eqn:6}
 J_{\rm C}(E) = 4.066 \, \frac{1}{\beta ^2} \, E ^{1.22} \left(\frac{E^{0.95} + 0.63^{0.95}}{1 + 0.63^{0.95}} \right) ^{-4.19},
\end{equation}
where the CR intensity $J(E)$ (given in particles\,m$^2$\,s$^{-1}$ \,sr$^{-1}$\,(GeV/nuc)$^{-1}$) is a function of kinetic energy per nucleon $E$ (given in GeV/nuc). These LIS's are summarized in Figure \ref{fig:12}.

We present the LIS's as approximated above as new LIS's for protons, Helium and Carbon, which should be of value for solar modulation modeling.

\section*{Acknowledgements}

The authors wish to thank the GALPROP developers and their funding bodies for access to and use of the GALPROP WebRun service. DB thanks the South African National Research Foundation (NRF) and the Centre for Space Research at the NWU for providing him with a bursary for his PhD studies. MSP acknowledges the financial support of the NRF under the Incentive and Competitive Funding for Rated Researchers, grants no: 87820 and 68198.


\begin{thebibliography}{18}

\ifx \bisbn   \undefined \def \bisbn  #1{ISBN #1}\fi
\ifx \binits  \undefined \def \binits#1{#1} \fi
\ifx \bauthor  \undefined \def \bauthor#1{#1} \fi
\ifx \batitle  \undefined \def \batitle#1{#1} \fi
\ifx \bjtitle  \undefined \def \bjtitle#1{#1}\fi
\ifx \bvolume  \undefined \def \bvolume#1{\textbf{#1}}\fi
\ifx \byear  \undefined \def \byear#1{#1} \fi
\ifx \bissue  \undefined \def \bissue#1{#1} \fi
\ifx \bfpage  \undefined \def \bfpage#1{#1} \fi
\ifx \blpage  \undefined \def \blpage #1{#1} \fi
\ifx \burl  \undefined \def \burl#1{\textsf{#1}} \fi
\ifx \doiurl  \undefined \def \doiurl#1{\textsf{#1}} \fi
\ifx \betal  \undefined \def \betal{\textit{et al.}} \fi
\ifx \binstitute  \undefined \def \binstitute#1{#1} \fi
\ifx \binstitutionaled  \undefined \def \binstitutionaled#1{#1} \fi
\ifx \bctitle  \undefined \def \bctitle#1{#1} \fi
\ifx \beditor  \undefined \def \beditor#1{#1} \fi
\ifx \bpublisher  \undefined \def \bpublisher#1{#1} \fi
\ifx \bbtitle  \undefined \def \bbtitle#1{#1} \fi
\ifx \bedition  \undefined \def \bedition#1{#1} \fi
\ifx \bseriesno  \undefined \def \bseriesno#1{#1} \fi
\ifx \blocation  \undefined \def \blocation#1{#1} \fi
\ifx \bsertitle  \undefined \def \bsertitle#1{#1} \fi
\ifx \bsnm \undefined \def \bsnm#1{#1} \fi
\ifx \bsuffix \undefined \def \bsuffix#1{#1} \fi
\ifx \bparticle \undefined \def \bparticle#1{#1} \fi
\ifx \barticle \undefined \def \barticle#1{#1} \fi
\ifx \bconfdate \undefined \def \bconfdate #1{#1} \fi
\ifx \botherref \undefined \def \botherref #1{#1} \fi
\ifx \url \undefined \def \url#1{\textsf{#1}} \fi
\ifx \bchapter \undefined \def \bchapter#1{#1} \fi
\ifx \bbook \undefined \def \bbook#1{#1} \fi
\ifx \bcomment \undefined \def \bcomment#1{#1} \fi
\ifx \oauthor \undefined \def \oauthor#1{#1} \fi
\ifx \citeauthoryear \undefined \def \citeauthoryear#1{#1} \fi
\ifx \endbibitem  \undefined \def \endbibitem {}\fi
\ifx \bconflocation  \undefined \def \bconflocation#1{#1} \fi
\ifx \arxivurl  \undefined \def \arxivurl#1{\textsf{#1}} \fi


\bibitem[\protect\citeauthoryear{{Adriani}
  et~al.}{2011}]{Pamela2011protonhelium}
\begin{barticle}
\bauthor{\bsnm{{Adriani}}, \binits{O.}},
\bauthor{\bsnm{{Barbarino}}, \binits{G.C.}},
\bauthor{\bsnm{{Bazilevskaya}}, \binits{G.A.}}, \betal:
\bjtitle{Science}
\bvolume{332},
\bfpage{69}
(\byear{2011})
\end{barticle}
\endbibitem

\bibitem[\protect\citeauthoryear{{Adriani}
  et~al.}{2014}]{Pamela2014boroncarbon}
\begin{barticle}
\bauthor{\bsnm{{Adriani}}, \binits{O.}},
\bauthor{\bsnm{{Barbarino}}, \binits{G.C.}},
\bauthor{\bsnm{{Bazilevskaya}}, \binits{G.A.}}, \betal:
\bjtitle{\apj}
\bvolume{791},
\bfpage{93}
(\byear{2014})
\end{barticle}
\endbibitem

\bibitem[\protect\citeauthoryear{{Bisschoff} and
  {Potgieter}}{2014}]{Bisschoff2014}
\begin{barticle}
\bauthor{\bsnm{{Bisschoff}}, \binits{D.}},
\bauthor{\bsnm{{Potgieter}}, \binits{M.S.}}:
\bjtitle{Astrophys. J.}
\bvolume{794},
\bfpage{166}
(\byear{2014})
\end{barticle}
\endbibitem

\bibitem[\protect\citeauthoryear{{Boezio}}{2014}]{Boezio2014}
\begin{barticle}
\bauthor{\bsnm{{Boezio}}, \binits{M.}}:
\bjtitle{Braz. J. Phys.}
\bvolume{44},
\bfpage{441}
(\byear{2014})
\end{barticle}
\endbibitem

\bibitem[\protect\citeauthoryear{{Herbst} et~al.}{2012}]{Herbst2012}
\begin{barticle}
\bauthor{\bsnm{{Herbst}}, \binits{K.}},
\bauthor{\bsnm{{Heber}}, \binits{B.}},
\bauthor{\bsnm{{Kopp}}, \binits{A.}},
\bauthor{\bsnm{{Sternal}}, \binits{O.}},
\bauthor{\bsnm{{Steinhilber}}, \binits{F.}}:
\bjtitle{\apj}
\bvolume{761},
\bfpage{17}
(\byear{2012})
\end{barticle}
\endbibitem

\bibitem[\protect\citeauthoryear{{Menn} et~al.}{2013}]{Menn2013}
\begin{barticle}
\bauthor{\bsnm{{Menn}}, \binits{W.}},
\bauthor{\bsnm{{Adriani}}, \binits{O.}},
\bauthor{\bsnm{{Barbarino}}, \binits{G.C.}}, \betal:
\bjtitle{Adv. Space Res.}
\bvolume{51},
\bfpage{209}
(\byear{2013})
\end{barticle}
\endbibitem

\bibitem[\protect\citeauthoryear{{Moskalenko}}{2011}]{Moskalenko}
\begin{botherref}
\oauthor{\bsnm{{Moskalenko}}, \binits{I.}}:
{arXiv:1105.4921v1 [astro-ph.HE]}
(2011)
\end{botherref}
\endbibitem

\bibitem[\protect\citeauthoryear{{Potgieter}}{2013}]{PotgieterReview2013}
\begin{barticle}
\bauthor{\bsnm{{Potgieter}}, \binits{M.S.}}:
\bjtitle{Living Rev. Solar Phys.}
\bvolume{10},
\bfpage{3}
(\byear{2013})
\end{barticle}
\endbibitem

\bibitem[\protect\citeauthoryear{{Potgieter}}{2014}]{Potgieter2014}
\begin{barticle}
\bauthor{\bsnm{{Potgieter}}, \binits{M.S.}}:
\bjtitle{Braz. J. Phys.}
\bvolume{44},
\bfpage{581}
(\byear{2014})
\end{barticle}
\endbibitem

\bibitem[\protect\citeauthoryear{{Potgieter} et~al.}{2014a}]{Potgieter2014Conf}
\begin{botherref}
\oauthor{\bsnm{{Potgieter}}, \binits{M.S.}},
\oauthor{\bsnm{{Vos}}, \binits{E.E.}},
\oauthor{\bsnm{{Nndanganeni}}, \binits{R.R.}}:
Proc. 14th ICATPP Conf. on Astroparticle, Particle, Space Physics and Detectors
  for Physics Applications, Como, Italy
(2014a)
\end{botherref}
\endbibitem

\bibitem[\protect\citeauthoryear{{Potgieter}
  et~al.}{2014b}]{Potgieter2014Solar}
\begin{barticle}
\bauthor{\bsnm{{Potgieter}}, \binits{M.S.}},
\bauthor{\bsnm{{Vos}}, \binits{E.E.}},
\bauthor{\bsnm{{Boezio}}, \binits{M.}},
\bauthor{\bsnm{{De Simone}}, \binits{N.}},
\bauthor{\bsnm{{Di Felice}}, \binits{V.}},
\bauthor{\bsnm{{Formato}}, \binits{V.}}:
\bjtitle{\solphys}
\bvolume{289},
\bfpage{391}
(\byear{2014}b)
\end{barticle}
\endbibitem

\bibitem[\protect\citeauthoryear{{Ptuskin} et~al.}{2006}]{Ptuskin2006}
\begin{barticle}
\bauthor{\bsnm{{Ptuskin}}, \binits{V.S.}},
\bauthor{\bsnm{{Moskalenko}}, \binits{I.V.}},
\bauthor{\bsnm{{Jones}}, \binits{F.C.}},
\bauthor{\bsnm{{Strong}}, \binits{A.W.}},
\bauthor{\bsnm{{Zirakashvili}}, \binits{V.N.}}:
\bjtitle{Astrophys. J.}
\bvolume{642},
\bfpage{902}
(\byear{2006})
\end{barticle}
\endbibitem

\bibitem[\protect\citeauthoryear{{Stone} et~al.}{2013}]{Stone2013}
\begin{barticle}
\bauthor{\bsnm{{Stone}}, \binits{E.C.}},
\bauthor{\bsnm{{Cummings}}, \binits{A.C.}},
\bauthor{\bsnm{{McDonald}}, \binits{F.B.}},
\bauthor{\bsnm{{Heikkila}}, \binits{B.C.}},
\bauthor{\bsnm{{Lal}}, \binits{N.}},
\bauthor{\bsnm{{Webber}}, \binits{W.R.}}:
\bjtitle{Science}
\bvolume{341},
\bfpage{150}
(\byear{2013})
\end{barticle}
\endbibitem

\bibitem[\protect\citeauthoryear{{Strauss} and
  {Potgieter}}{2014}]{StraussPotgieter2014}
\begin{barticle}
\bauthor{\bsnm{{Strauss}}, \binits{R.D.}},
\bauthor{\bsnm{{Potgieter}}, \binits{M.S.}}:
\bjtitle{Adv. Space Res.}
\bvolume{53},
\bfpage{1015}
(\byear{2014})
\end{barticle}
\endbibitem

\bibitem[\protect\citeauthoryear{{Strong} and
  {Moskalenko}}{1998}]{StrongMoskalenko1998}
\begin{barticle}
\bauthor{\bsnm{{Strong}}, \binits{A.W.}},
\bauthor{\bsnm{{Moskalenko}}, \binits{I.V.}}:
\bjtitle{Astrophys. J.}
\bvolume{509},
\bfpage{212}
(\byear{1998})
\end{barticle}
\endbibitem

\bibitem[\protect\citeauthoryear{{Strong} et~al.}{2007}]{StrongReview2007}
\begin{barticle}
\bauthor{\bsnm{{Strong}}, \binits{A.W.}},
\bauthor{\bsnm{{Moskalenko}}, \binits{I.V.}},
\bauthor{\bsnm{{Ptuskin}}, \binits{V.S.}}:
\bjtitle{Ann. Rev. Nucl. Part. Sci.}
\bvolume{57},
\bfpage{285}
(\byear{2007})
\end{barticle}
\endbibitem

\bibitem[\protect\citeauthoryear{{Vladimirov} et~al.}{2011}]{Webrun}
\begin{barticle}
\bauthor{\bsnm{{Vladimirov}}, \binits{A.E.}},
\bauthor{\bsnm{{Digel}}, \binits{S.W.}},
\bauthor{\bsnm{{J{\'o}hannesson}}, \binits{G.}},
\bauthor{\bsnm{{Michelson}}, \binits{P.F.}},
\bauthor{\bsnm{{Moskalenko}}, \binits{I.V.}},
\bauthor{\bsnm{{Nolan}}, \binits{P.L.}},
\bauthor{\bsnm{{Orlando}}, \binits{E.}},
\bauthor{\bsnm{{Porter}}, \binits{T.A.}},
\bauthor{\bsnm{{Strong}}, \binits{A.W.}}:
\bjtitle{Comp. Phys. Comm}
\bvolume{182},
\bfpage{1156}
(\byear{2011})
\end{barticle}
\endbibitem

\bibitem[\protect\citeauthoryear{{Vos} et~al.}{2013}]{Vos2013}
\begin{botherref}
\oauthor{\bsnm{{Vos}}, \binits{E.E.}},
\oauthor{\bsnm{{Potgieter}}, \binits{M.S.}},
\oauthor{\bsnm{{Boezio}}, \binits{M.}},
\oauthor{\bsnm{{Di Felice}}, \binits{V.}},
\oauthor{\bsnm{{De Simone}}, \binits{N.}},
\oauthor{\bsnm{{Formato}}, \binits{V.}},
\oauthor{\bsnm{{Munini}}, \binits{R.}}:
Proc. 33rd Int. Cosmic Ray Conf., Rio de Janeiro, Brazil
(2013)
\end{botherref}
\endbibitem

\end{thebibliography}

\newpage

\end{document}